\documentclass[reprint,aps,prl,superscriptaddress,showpacs]{revtex4-2}

\usepackage[utf8]{inputenc}
\usepackage{graphicx}
\usepackage{tikz}
\usepackage{dcolumn}
\usepackage{bm}
\usepackage{amsmath}
\usepackage{amsfonts}
\usepackage{amssymb}
\usepackage{hyperref}
\usepackage{pbox}
\usepackage{pinlabel}
\usepackage{ragged2e}
\usepackage[justification=RaggedRight]{caption}
\usepackage[singlelinecheck=off]{subcaption}
\usepackage{ulem}

\begin{document}
\title{Stabilization of Fermi-liquid behavior by interactions in disordered metals}

\author{Arianna Poli}	
\affiliation{Dipartimento di Scienze Fisiche e Chimiche, Università dell’Aquila, Coppito-L’Aquila, Italy}	
\author{Simone Fratini}
\affiliation{Universit\'e Grenoble Alpes, CNRS, Grenoble INP, Institut N\'eel, 38000 Grenoble, France}
\author{Jennifer Coulter}
\affiliation{Center for Computational Quantum Physics, The Flatiron Institute,162 5th Avenue,New York, NY 10010}
\author{Andrew J. Millis}
\affiliation{Center for Computational Quantum Physics, The Flatiron Institute,162 5th Avenue,New York, NY 10010}
\affiliation{Department of Physics, Columbia University, 538 West 120th Street, New York, New York 10027}
\author{Sergio Ciuchi}%
\affiliation{Dipartimento di Scienze Fisiche e Chimiche, Università dell’Aquila, Coppito-L’Aquila, Italy}

\begin{abstract}
We study the interplay of electron-electron and electron-disorder scattering in correlated Fermi liquids by the disordered Hubbard model using dynamical mean-field theory with an IPT-CPA solver. We find significant violations of Matthiessen's rule (additivity of scattering mechanisms) which we explain in terms of the screening of the disorder potential by interactions, leading to a protection of the electron-electron inelastic scattering rate against disorder. We also show that large disorder can lead to a surprising enhancement of the electron-electron scattering that contrasts with the competition seen in the elastic channel.  Our results compare positively with available resistivity data in the disordered Fermi liquid phase of the correlated organic metals $\kappa$-(ET)$_{2}$X, and rationalize the strong sample dependence of the $T^2$ coefficients observed in the resistivity of correlated perovskite oxides such as SrVO$_3$.
\end{abstract}

\maketitle

\paragraph{Introduction.---}

The interplay between electronic interactions and disorder is a fundamental problem in condensed matter research \cite{Lee-RMP85}, revealing complex phases and unusual transport properties of the electron gas both in one \cite{Giamarchi-book1D} and two dimensions \cite{Punnoose-Science05,Finkelstein-book,Abrahams-RMP01}. This interplay is also generally relevant to the metal-insulator transition  in strongly correlated materials, where randomness is not only inevitably present in laboratory conditions, but can  be experimentally controlled. In this respect, one particularly interesting and experimentally well studied system is the $\kappa$-organics, a prototypical class of half-filled Mott systems, where a remarkable sensitivity to hydrostatic pressure has revealed multiple phases within a relatively accessible pressure range, enabling a careful study of the physical properties across a 1st order Mott transition \cite{Dressel_NMat,Dressel_NComm,PhysRevLett.91.016401_phaseDiagram_disU_limelette} and where a high sensitivity to disorder, introduced through x-ray or proton irradiation, is observed \cite{Analytis-kOrganicsPRL2006,Urai-Kanoda-PRB19,Sasaki-JPSJ07,Sasaki-PRL08}.

On the theoretical side, it is understood that while both strong local interactions and disorder can separately drive metal-insulator transitions in half-filled systems (of the Mott and Anderson type respectively), these two mechanisms compete rather than cooperate: while the latter naturally drives spatial inhomogeneities of the charge density, the former acts to make the density homogeneous by quenching the charge fluctuations. 

The 
consequences of this competition taking place at the microscopic level \cite{Tanaskovic_HubbardAndersonKondoScreening_PRL2003} are best seen from the demonstrated tendency for disorder to stabilize the metallic phase, shifting the Mott transition to larger values of the interaction strength \cite{Byczuk-Vollhardt-PRL05,Aguiar_Mott_typical_medium_PhysRevLett.102.156402,RadonjicPRB2010}, in agreement with the experimental observations \cite{Urai-Kanoda-PRB19,Sasaki-JPSJ07,Sasaki-PRL08,Analytis-kOrganicsPRL2006}. 

The effects of this interplay on the high-temperature transport properties have also been analyzed theoretically \cite{RadonjicPRB2010,CiuchiFratiniStrangeMetal}, 
and specific results have been obtained down to low temperatures \cite{Aguiar-EPL04}.
Yet, obtaining a systematic understanding of the interplay between disorder-induced and correlation-induced electron scattering in the low-temperature Fermi liquid regime remains an open problem.  

Here we demonstrate that low-temperature transport is also characterized by a marked breakdown of Matthiessen's rule (additivity of scattering from different mechanisms). We find a suppression of the effective disorder strength by electronic correlations which implies that the Fermi liquid properties, embodied in the quadratic temperature dependence of the scattering rate arising from electron-electron interactions, become increasingly insensitive to disorder with increasing interactions. We also find that this protection breaks down beyond a certain threshold, where disorder starts instead to cooperate with interactions, leading to an enhancement of the quadratic temperature coefficient.

\paragraph{Model and method.---} 
We study the following disordered Hubbard model defined by:
    \begin{equation}
        H=-t\sum_{i,j,\sigma} c^{\dagger}_{i,\sigma}c_{j,\sigma} +U\sum_{i}n_{i,\uparrow}n_{i,\downarrow} -\sum_i\xi_i c^{\dagger}_{i,\sigma}c_{i,\sigma}
    \end{equation}
where $c^{\dagger}_{i,\sigma}$ and $c_{i,\sigma}$ are  fermionic creation and annihilation operators,  $n_{i,\sigma}=c^{\dagger}_{i,\sigma}c_{i,\sigma}$ the local densities, $t$ is the hopping term, $U$ is a local electron-electron repulsion term. $\xi_i$ is a zero mean site-uncorrelated Gaussian variables of variance $W^2$. 
We assume a semi-circular band density of states $\mathcal{D}(\varepsilon)=\frac{2}{\pi D^2}\sqrt{D^2-\varepsilon^2}$ and measure all energies in units of the half-bandwidth $D$. Typical values in organic salts are $D\simeq 0.25$eV  \cite{Kandpal-PRL09}. 

Combining interactions and disorder is a notoriously difficult problem, with only few reliable theoretical methods available. Here we employ single-site Dynamical Mean Field Theory (DMFT) to address correlations, utilizing the Iterated Perturbation Theory (IPT) impurity solver and combine it with the Coherent Potential Approximation (CPA) for disorder \cite{Aguiar-EPL04,Aguiar_PhysRevB.71.205115,PhysRevB.106.195156_Fotso_DMFT_CPA_nonequil,Mooij}, achieving reliable results in the correlated metallic regime of interest where Anderson localization corrections can be neglected.
We use the method of Ref. \cite{Aguiar-EPL04}, but employing the interpolation scheme of \cite{Pothoff_Theory_PhysRevB.55.16132} to deal with disorder-induced local deviations from half-filling. The algorithm is detailed in Supplementary Sec. I.  
Using CPA an averaged impurity model Green function is used in the DMFT self consistency condition, the resulting self energy $\Sigma(\omega)$ now includes
effects both of disorder and of interactions. The resistivity is then obtained by standard methods from the disorder averaged spectral function via the Kubo formula  
using appropriate current vertex \cite{Chattopadhyay-OptSW-PRB2000} (see Supplementary sec. II).

\paragraph{Resistivity and scattering rate.---}
\begin{figure}[h]
\centering
\includegraphics[width=0.4\textwidth]{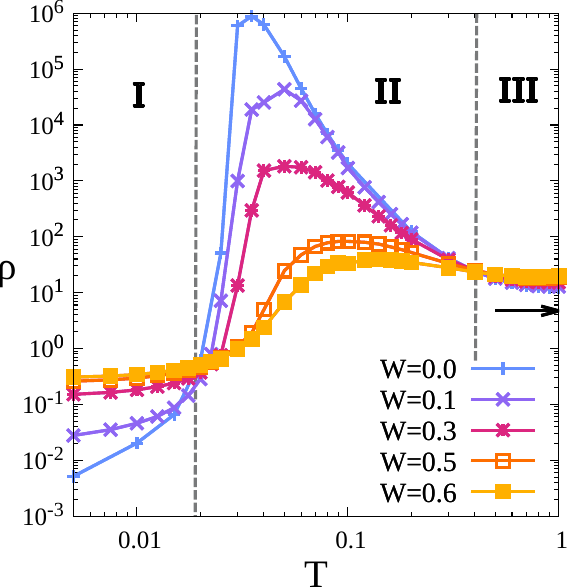}
\caption{Resistivity calculated for  different values of disorder at $U/D=3.0$, a value close to the MIT at $W=0$ ($U_c=3.2$). 
The unit of resistivity is $\bar\rho=(v/a^2)\hbar/e^2$, with $v$ the unit cell volume, $a$ the in-plane lattice parameter and $e$ the electron charge. This yields $\bar\rho=266\mu\Omega$cm on a triangular lattice with typical interlayer spacing $d=15\AA$. The arrow marks the MIR limit (see text). 
}
\label{fig:resistivityall}
\hfill
\end{figure}
Fig. \ref{fig:resistivityall} shows the resistivity calculated in the strongly correlated metallic phase for  $U/D=3.0$ near the  metal-insulator transition (MIT) of the clean system ($U_{c2}/D=3.2$ within IPT) for different levels of disorder. 
Three different regimes can be distinguished for the resistivity, corresponding to  low, intermediate  and high temperatures. These can be identified respectively as: (I) normal disordered metal, (II) anomalous and strongly disorder-dependent transport, above the crossing point (here at $T/D\simeq 0.02$) and (III) incoherent transport, weakly disorder dependent.

The anomalous regime (II) at intermediate temperatures has been explored in previous works \cite{CiuchiFratiniStrangeMetal,RadonjicPRB2010}. Here the fact  that the metal-insulator transition is shifted to larger $U$ by increasing $W$  \cite{Aguiar_PhysRevB.71.205115,Aguiar_Mott_typical_medium_PhysRevLett.102.156402}
leads to the surprising result that the resistivity decreases with increasing randomness. This effect characterizes the correlated metallic phase at intermediate temperatures for a wide range of $U$ 
and it becomes particularly strong (and it occurs on a wide temperature window) in proximity of the Mott transition, as shown in Fig. \ref{fig:resistivityall}. 
In the high-temperature regime (III) the resistivity is instead insensitive to the locus of the zero-temperature phase transition. It increases weakly as the disorder strength increases and at high T saturates to values moderately exceeding the Mott-Ioffe-Regel limit,  $\rho_{MIR}=3\pi/2$ in the present units \cite{CiuchiFratiniStrangeMetal}.

Our main focus here is on the less-explored low-temperature regime (I) below the crossing point. Here an apparently normal behavior is recovered, where  the resistivity is an increasing function of disorder strength and a nonzero residual resistivity appears at $T=0$, as expected for disordered metals \cite{AshcroftMermin76}.

\begin{figure}[h]
\centering
\includegraphics[width=0.4\textwidth]{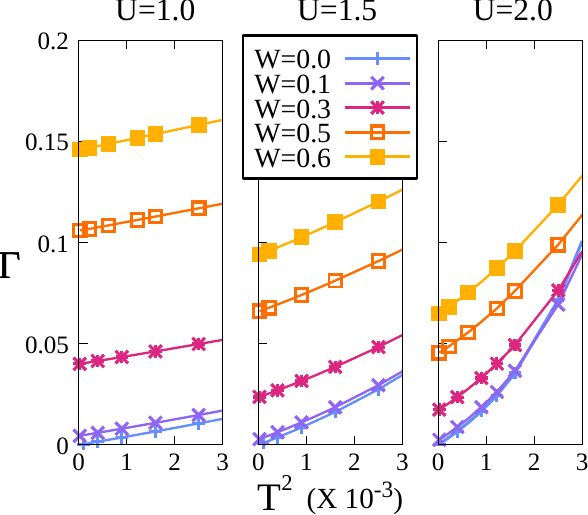}
\caption{Scattering rate as a function of $T^2$ for $U$-values indicated. At fixed $U$, the intercept varies but the slope at low $T$ remains almost independent of disorder mirroring Fig. 3 in Ref. \cite{Urai-Kanoda-PRB19} 
}
\label{fig:GammavsT2}
\hfill
\end{figure} 

At low temperatures and for weak to moderate disorder strengths, the resistivity is proportional to the scattering rate at the Fermi energy, $\Gamma=-Im\Sigma(\omega=0)$ (see Refs. \cite{Millis99,Mooij} and Supplementary Sec. II). Fermi liquid theory predicts that the latter obeys the general expression:
\begin{equation}
    \Gamma=\Gamma_0+bT^2,
    \label{eq:GammaLowTemp}
\end{equation} 
where $\Gamma_0$ is an elastic term originating from disorder scattering and $bT^2$ is an inelastic term coming from electron-electron collisions. 
As shown in Fig. (\ref{fig:GammavsT2}), this form applies for temperatures less than a $U$-dependent scale. At all values of $U$, the intercept $\Gamma_0$ markedly increases with the disorder strength,  while the slope $b$ is much less affected by disorder, in agreement with available experiments \cite{Urai-Kanoda-PRB19,Sasaki-JPSJ07,Sasaki-PRL08,Analytis-kOrganicsPRL2006} (see below for a more detailed comparison).
The competition between randomness and interactions, which causes a disorder-induced metallization at large temperatures, is seen here as a decrease of $\Gamma_0$ upon increasing $U$ for any given level of disorder. We next quantitatively analyze and explain these observations.

\paragraph{Renormalized disorder and interactions.---}
The standard picture of systems with more than one type of scattering is given by Matthiessen's rule \cite{AshcroftMermin76}, stating the additivity of scattering rates. In the context of Eq. (\ref{eq:GammaLowTemp}) Matthiessen's rule is $\Gamma(U,W)=\Gamma(U=0,W)+\Gamma(U,W=0)$ implying $b$ is independent of the disorder strength and $\Gamma_0$ is independent of the interaction strength. Comparison of the three panels of Fig.~\ref{fig:GammavsT2} shows immediately that Matthiessen's rule does not hold for the disordered Hubbard model since $\Gamma(T=0)$ depends on interaction at fixed $W$.

\begin{figure}[h]
\centering
\includegraphics[width=0.35\textwidth]{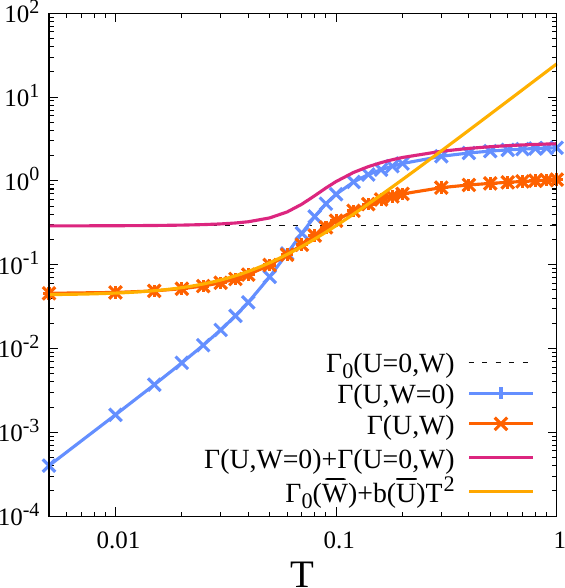}
\caption{Scattering rate function for $U=2.0$ and $W=0.5$ calculated within the IPT+CPA method (orange, $\times$) compared to: the bare interaction scattering rate (light blue, $+$), bare Matthiessen's rule calculation (purple), and the sum of residual scattering rate using  renormalized disorder (yellow). Horizontal dashed line shows the temperature independent scattering rate in the non-interacting disordered case.
\label{fig:Matthiessen}
}
\end{figure}

In Fig. ~\ref{fig:Matthiessen} we investigate the additivity of scattering rates in more detail for one specific choice of interaction and disorder. The blue trace is the scattering rate calculated for $W=0$ and $U\neq 0$; the Fermi liquid $T^2$ low T behavior rolling over to $T$-independent at high T is evident. The value of the temperature-independent scattering rate expected at $U=0$ is shown as the black arrow. The sum of the two rates is shown as a purple line in Fig. (\ref{fig:Matthiessen}). The actual scattering rate calculated in the presence of both $U$ and $W$ (orange, $\times$ symbols) is almost an order of magnitude lower than the Matthiessen result and has a modestly different temperature dependence.
This discrepancy is not only substantial but also opposite to the general rule that sets the additive result as a lower bound to the actual scattering rate \cite{Ziman_el_ph_book}.

Specializing  to the disordered Fermi liquid regime at low temperatures, the observations of the preceding paragraphs imply that we can still parametrize the scattering rate with Eq. (\ref{eq:GammaLowTemp}). For this we need to introduce parameters $\Gamma_0$ and $b$ that 
are not independent functions of disorder and interactions, but depend instead on both $U$ and $W$.

\begin{figure*}[ht]
\includegraphics[width=0.95\textwidth]{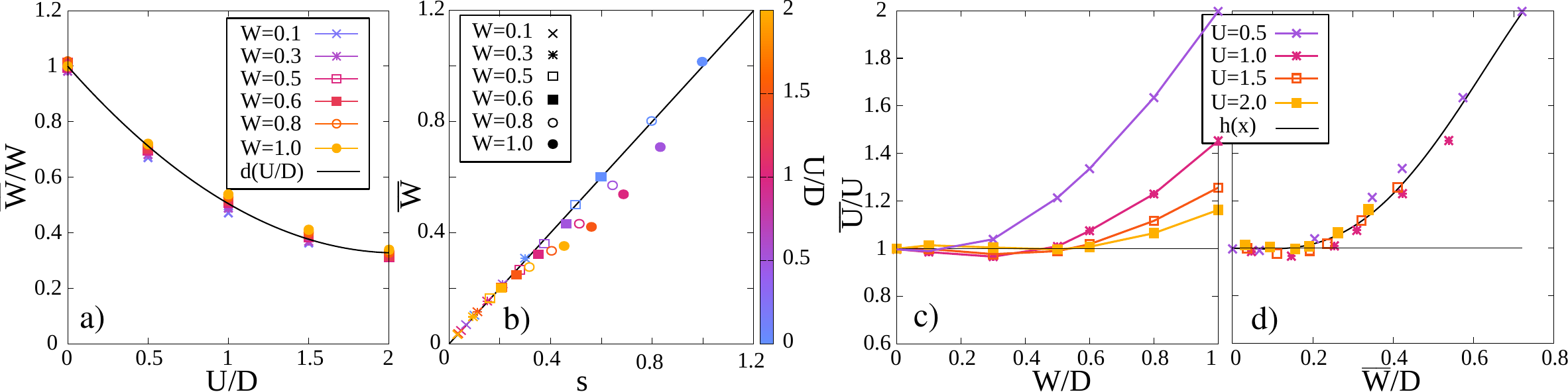}

\caption{a) The renormalized disorder parameter $\Bar{W}/W$ as a function of $U$ for different values of disorder $W$, black line is a fitting curve $d(x)=1+d_1 x+d_2 x^2$. b) Renormalized disorder parameter versus the standard deviation $s=\sqrt{\langle \eta^2 \rangle}$ of the screened random potential. c) Effective interaction $\Bar{U}/U$ as a function of the disorder. d) Effective interaction parameter $\Bar{U}/U$ as a function of the renormalized disorder parameter $\bar{W}/D$, \ black line is a fitting curve $h(x)=1+h_2 x^2+h_3 x^3+h_4 x^4$. 
}
\label{fig:Wbar_Ubar}
\hfill
\end{figure*} 

The result of a quadratic fit of the data through Eq. (\ref{eq:GammaLowTemp}) is shown as a solid yellow line in Fig. (\ref{fig:Matthiessen}) for the particular case $U/D=2$, $W/D=0.5$. We can now introduce the renormalized parameters $\Bar{U}$ and $\Bar{W}$ that would describe the numerical results in the spirit of the additive rule. First, for each pair $(U,W)$ we define $\Bar{W}$ as the value of the disorder strength that would yield the calculated $\Gamma_0$ if the system was non-interacting, namely $\Gamma_0(U=0,\bar{W})\equiv \Gamma_0(U,W)$ (the functional form of $\Gamma_0(U=0,\bar{W})$ is known analytically within the CPA and can be easily inverted to yield  $\bar{W}$ (see Supplementary Sec. V).
Second, we define a renormalized interaction $\bar{U}$ as the pure-system $U$ that would reproduce the value of $b$ calculated for the disordered system:  $b(\bar{U},W=0)\equiv b(U,W)$. In practice we obtain  $b$ from  a linear fitting of  $\Gamma$ vs $T^2$ at low temperature ($T^2<8\cdot10^{-4}$, see Fig. \ref{fig:GammavsT2}). The function $b(U,W=0)$ needed to obtain $\bar{U}$  is extracted from the data using the polynomial expression $a_0 U^2+a_4 U^4$ with $a_0$ known from the perturbation expansion. Obviously, the renormalized parameters $\Bar{U}$ and $\Bar{W}$ are now functions of both $U$ and $W$. The results are reported in Fig. (\ref{fig:Wbar_Ubar}), summarizing the main conclusions of our work.

Fig. \ref{fig:Wbar_Ubar}(a) shows that the effects of disorder are suppressed by interactions, and that this suppression happens as soon as $U>0$.
Moreover, all curves for $\Bar{W}/W$ collapse onto a universal curve when plotted as a function of $U$ alone: this means that disorder is effectively suppressed by interactions in the same proportions, whether it is weak or strong. 

We can understand the observed reduction of the disorder strength as arising from the density response of the interacting electron system\cite{Mooij,DiSanteCiuchi_Strong_interplay_PhysRevB.90.075111,Tanaskovic_HubbardAndersonKondoScreening_PRL2003}. An on-site disorder potential $\xi$ changes the local electron density, which in turn induces a Hartree potential that tends to partially compensate it.
The random potential seen by the electrons is therefore $\eta=\xi+\nu$, where the term $\nu$ originating from the  interaction self-energy is  opposite to the driving term $\xi$ (see Supplementary Sec. IV).
As a result, the variance of the screened potential is always $s^2\equiv \langle\eta^2\rangle < W^2$. 
As shown in Fig. \ref{fig:Wbar_Ubar}(b) $\bar{W}$ defined previously from the residual scattering rate $\Gamma_0$ closely follows the standard deviation of the screened potential. 
$\Gamma_0$ itself approximately scales with $s^2$ for small $s/D$ (see Supplementary Sec. IV). 

Fig. \ref{fig:Wbar_Ubar}(c) reports  the renormalized interaction strength, $\Bar{U}/U$, as a function of disorder. The data indicate that the interaction contribution to scattering is initially protected,  staying pinned to the clean Fermi liquid limit up to a threshold value of $W$ that increases with $U$. Beyond this threshold, disorder leads to an \textit{increase} of the effective interaction strength, in apparent contradiction with the competition found in all other quantities. 
This behavior is more clearly illustrated  in Fig.~\ref{fig:Wbar_Ubar}(d), where the renormalized interaction is shown to exhibit approximate scaling when plotted in terms of the renormalized disorder. The protection persists up to $\bar{W}/D \simeq 0.2$, meaning it is determined by the strength of the screened disorder rather than its bare value. Beyond this threshold, the protection ceases, and the effective interaction increases as the screened disorder strengthens.

\paragraph{Comparison to experiments.---}
\begin{figure}[h]
\centering
\includegraphics[width=0.45\textwidth]{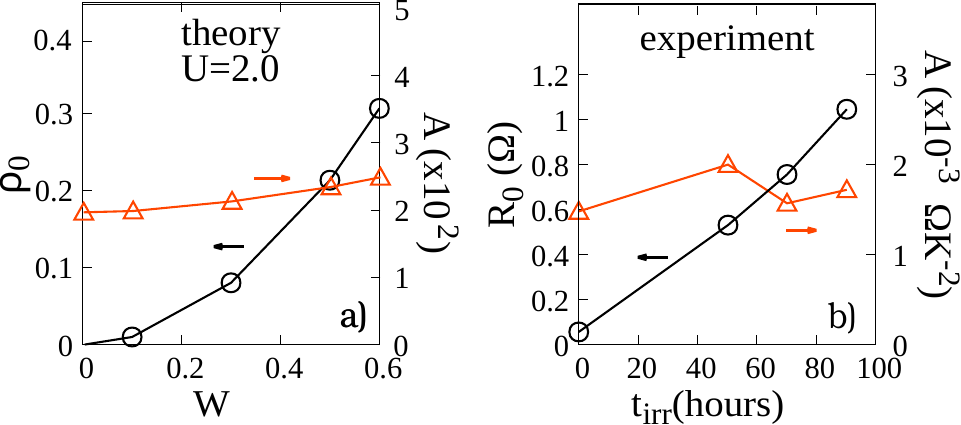}
\caption{a)  Coefficients of the resistivity $\rho(T)=\rho_0+AT^2$  as a function of disorder calculated in the disordered Hubbard model for $U/D=2$. 
b) Analogous quantities measured in the $\kappa$-ET$_2$X organics as a function of X-ray irradiation, reproduced from Ref. \cite{Urai-Kanoda-PRB19}.
}
\label{fig:R_0_A}
\hfill
\end{figure} 
The overall behavior reported in Fig. \ref{fig:resistivityall}, exhibiting three different transport regimes, matches what is observed in the $\kappa$-ET$_2$X organics \cite{Analytis-kOrganicsPRL2006}. Similar to our theoretical results, broad violations of Matthiessen's rule have been found experimentally.
For a more in-depth comparison with our findings, we specifically analyze the X-ray irradiation and pressure dependence of the Fermi-liquid behavior that was reported in Ref. \cite{Urai-Kanoda-PRB19}. 
In the experimental system, disorder was induced by X-ray irradiation, while the bandwidth, and thus the effective interaction  strength $U/D$  is tuned by pressure. We present an analysis of data obtained as a function of irradiation at fixed pressure. The pressure dependence requires a more involved analysis, given in Section VI of the Supplementary material and summarized here.

The measured resistance at all pressures and X-ray irradiation times is found to have the general expression $R=R_0+AT^2$, consistent with Eq. (\ref{eq:GammaLowTemp}). Fig.~\ref{fig:R_0_A} compares the zero temperature resistance and $T^2$ coefficent theoretically calculated as a function of disorder strength $W$ to the same quantities experimentally measured in Ref. \cite{Urai-Kanoda-PRB19}. In the experiment $T\rightarrow 0$ value of the resistance is found to increase with irradiation time, while the coefficient of the $T^2$ term is found to depend only weakly on irradiation time, consistent with our calculations. The theory results are similarly consistent with those of Ref. \cite{Analytis-kOrganicsPRL2006}.  Further, while the data points are relatively sparse, careful examination of the experimental $R_0$ vs irradiation time data reveals a nonlinearity: a straight-line extrapolation of the measured values at the two highest $t_{irr}$ to $t_{irr}=0$ has a negative intercept; again qualitatively consistent with our calculation.

In addition to the effect of disorder, Ref. \cite{Urai-Kanoda-PRB19} has also studied the effect of hydrostatic pressure for a fixed level of disorder, finding that residual resistance $R_0$ decreases markedly with pressure. A theoretical description of the pressure dependence is more complicated because the pressure increases the bandwidth, simultaneously decreasing the relative disorder strength $W/D$ and weakening the interaction $U/D$, which has the effect of descreasing the interaction-induced screening of the disorder and thus partially compensating the effect $W/D$ by increasing the effective disorder strength (see Supplementary VI).  

Our theory also predicts a decrease of the temperature coefficient $A$ with increasing pressure due to the decrease of $U/D$, which is also observed in the experiment. Finally, the  temperature marking the end of Fermi liquid behavior increases with pressure, which can be also ascribed to a reduced $U/D$. We conclude that our model calculations provide a consistent description of the disordered Fermi liquid phase of the $\kappa$-ET$_2$X organics.

\paragraph{Conclusions---.}
Previous theoretical works have reported how the interplay of interactions and disorder manifests in the high-temperature transport properties of disordered correlated systems. 
The analysis presented here demonstrates how this interplay also affects the low-temperature Fermi-liquid behavior, $\rho=\rho_0+AT^2$. 

Screening of the random potentials by interactions causes a reduction of the residual resistivity $\rho_0$ as compared with non-interacting systems. The effects are instead more complex regarding the quadratic temperature coefficient that embodies electron-electron collisions. On one hand, by screening the effective disorder potential, large values of $U$ appear to stabilize Fermi liquid behavior \cite{Tanaskovic_HubbardAndersonKondoScreening_PRL2003}, pinning the coefficient $A$ to its disorder-free value. On the other hand, this protection against disorder breaks down when the latter becomes too strong: beyond a certain threshold, disorder starts to cooperate with interactions causing an \textit{increase} of the inelastic scattering rate, which can be interpreted as an increase of the effective interaction strength.

Our observations closely match the available experiments in $\kappa$-ET$_2$X organic salts, a class of one-orbital correlated systems where magnetic order is frustrated, granting access to the low-temperature Fermi liquid phase, and where the strength of both correlations and disorder can be 
controlled by hydrostatic pressure and irradiation. More generally, our results could provide an explanation to the large sample-to-sample spread of the $T^2$ resistivity coefficients observed in moderately correlated perovskite metals such as SrVO$_3$ \cite{Lee-Hand-arxiv24}: these large variations of $A$ are accompanied by (and correlate with) even larger variations of the residual resistivity $\rho_0$, indicative of a dominant role of extrinsic disorder.  

In the future it will be interesting to explore how the competition of disorder and interactions manifests in the "strange" metallic state  reached at large $U$ by doping Mott systems away from half filling \cite{Rullier-albenque1,Rullier-albenque2,Rullier-albenque3} as well as in more complex, multi-orbital \cite{Vlad-Mooij-NanoLett2021}  and Hund materials \cite{Abramovitch-ee-eph-PRL2024}. Finally, the ideas put forward here could be used to revisit the competition between electronic correlations and  electron-phonon interactions, bringing them under the present lens: thermal phonons behave as a temperature-dependent, intrinsic source of disorder \cite{Millis99,Mooij},  that could also lead to an effective increase of the electron-electron interaction strength.
\acknowledgments
A.P. acknowledges Olivier Parcollet and Herbert Fotso for useful discussions.
S.C. acknowledges funding from  NextGenerationEU National Innovation Ecosystem grant ECS00000041 - VITALITY - CUP E13C22001060006 and grant PE00000023 - IEXSMA - CUP E63C22002180006. The work of A.P. was supported in part by the Simons foundation via a Flatiron Institute predoctoral fellowship. The Flatiron Institute is a division of the Simons foundation.

\bibliography{references.bib}

\end{document}